\documentclass[12pt,a4paper]{article}
\usepackage{geometry}
\usepackage[ngerman, english]{babel}
\usepackage{hyperref}
\usepackage{xr-hyper}
\usepackage{amsmath,amsfonts,amssymb}
\usepackage{url}
\usepackage{graphicx}
\usepackage{caption}
\usepackage{authblk}
\usepackage{subfig}
\usepackage{comment}
\usepackage[dvipsnames]{xcolor}
\usepackage[normalem]{ulem}

\title{Multi-mode lasing in supercell plasmonic nanoparticle arrays}
\author[1]{Rebecca Heilmann}
\author[1]{Kristian Arjas}
\author[2]{Tommi K. Hakala}
\author[1,*]{P{\"a}ivi T{\"o}rm{\"a}}
\affil[1]{Department of Applied Physics, Aalto University School of Science, P.O. Box 15100, Aalto, FI-00076, Finland}
\affil[2] {Center for Photonics Sciences, University of Eastern Finland, P.O. Box 111, FI-80101 Joensuu, Finland}
\affil[*]{paivi.torma@aalto.fi}

\date{}

\makeatletter
\newcommand*{\addFileDependency}[1]{
  \typeout{(#1)}
  \@addtofilelist{#1}
  \IfFileExists{#1}{}{\typeout{No file #1.}}
}

\makeatother

\newcommand*{\myexternaldocument}[1]{
    \externaldocument{#1}
    \addFileDependency{#1.tex}
    \addFileDependency{#1.aux}
}
\myexternaldocument{SI}

\begin{document}

\maketitle
\clearpage
\begin{abstract}
Multicolour light sources can be used in applications such as lighting and multiplexing signals. In photonic and plasmonic systems, one way to achieve multicolour light is via multi-mode lasing. To achieve this, plasmonic nanoparticle arrays are typically arranged in superlattices that lead to multiple dispersions of the single arrays coupled via the superlattice Bragg modes. Here, we show an alternative way to enable multi-mode lasing in plasmonic nanoparticle arrays. We design a supercell in a square lattice by leaving part of the lattice sites empty. This results in multiple dispersive branches caused by the supercell period and hence creates additional band edges that can support lasing. We experimentally demonstrate multi-mode lasing in such a supercell array. Furthermore, we identify the lasing modes by calculating the dispersion by combining the structure factor of the array design with an empty lattice approximation. We conclude that the lasing modes are the 74th $\Gamma$- and 106th $X$-point of the supercell. By tuning the square lattice period with respect to the gain emission we can control the modes that lase. Finally, we show that the lasing modes exhibit a combination of transverse electric and transverse magnetic mode characteristics in polarization resolved measurements. 
\end{abstract}
\textbf{keywords:} plasmonics, nanophotonics, surface plasmon resonance, multi-mode lasing

\section*{Introduction}
Plasmonic nanoparticle arrays support surface lattice resonances (SLRs) that are dispersive plasmonic-photonic modes arising from a hybridization between the plasmonic resonances of individual nanoparticles and the diffracted orders governed by the lattice geometry. The spectral position of the SLRs can be easily tailored by varying the lattice geometry and period while simultaneously yielding high quality (Q-) factors~\cite{kravets2008extremely,chu2008experimental,auguie2008collective,giannini2011plasmonic,rodriguez2011coupling,humphrey2014plasmonic,ross2016optical}. Combined with emitters such as organic dye molecules, plasmonic nanoparticle arrays are an effective system to study light-matter interaction such as strong coupling or Bose-Einstein condensation~\cite{vakevainen2014plasmonic,bozhevolnyi2017quantum,hakala2018bose,de2018interaction,wang2018rich,vakevainen2020sub}. Lasing in plasmonic nanoparticle arrays has been studied in various lattice geometries such as square, rectangular, honeycomb or hexagonal lattices. Typically the systems produce lasing at a band edge originating at high symmetry points of the lattice, for instance at the $\Gamma$-, $K$- or $M$-points~\cite{zhou2013lasing,schokker2014lasing,yang2015real,ramezani2017plasmon,guo2019lasing, juarez2022m}. 
Also bound states in continuum which have extraordinary high Q-factors have been recently exploited for lasing in plasmonic arrays~\cite{heilmann2022quasi,mohamed2022controlling,salerno2022loss}.\\
For lighting applications and optical communication, multicolour light sources are necessary. Ideally, such sources span red, blue and green wavelengths in order to create white light or NIR regions for optical communication~\cite{lu2014all,liu2018laser,zhao2019full,guan2021plasmonic}. 
In photonic systems, one way to achieve multicolour light sources is via multi-mode lasing,  i.e. simultaneous lasing at a set of different modes. As lasing occurs in plasmonic systems at band edges, multiple band edges need to be created to realize multi-mode lasing. The most straightforward approach is by organizing individual arrays in a larger superlattice network, where the SLRs of the individual arrays couple to the Bragg modes of the superlattice, leading to multiple band edges at different energies and wavevectors~\cite{wang2015superlattice}. Multi-mode lasing has been observed in such superlattice geometries~\cite{wang2017band,wang2019manipulating}. Another possibility to realize multimode lasing is by dividing a square array into smaller patches which have slightly different periods. This creates additional band edges at different energies which simultaneously lase under optical pumping \cite{guo2019plasmon}.\\
Another way of creating additional band edges for a zero wavevector, i.e. into the direction normal to the array plane, is by introducing an effective second lattice period which yields a second SLR. This can be done in bipartite arrays~\cite{humphrey2014plasmonic,cuartero2020super,lim2022fourier} or by introducing periodic vacancies to the arrays~\cite{zundel2021lattice}.  By removing particles at designated positions, deterministic aperiodic lattices that yield more complicated band structures and hence additional band edges have been realised and lasing in such lattices has been demonstrated~\cite{schokker2016lasing}. However, multi-mode lasing has not been explicitly studied. Other systems in which multi-mode lasing has been observed include low-symmetry arrays, where two polarization dependent modes lase simultaneously~\cite{knudson2019polarization}, light-cone SLRs overlapping with higher Brillouin Zone (BZ) edges enabling lasing from several high symmetry points at once~\cite{guan2021identification}, and lasing in quasi-propagating modes that span a continuum of energies~\cite{tan2022lasing}. 
In addition to the aforementioned plasmonic structures, multi-mode lasing has been achieved in various photonic systems such as hyperuniform structures ~\cite{lin2020chip}, topological insulators~\cite{kim2020multipolar,gong2020topological}, bound states in the continuum~\cite{zhai2023multimode,mohamed2022controlling} and as simultaneous lasing in the magnetic and electric resonances of a dielectric nanoparticle array \cite{azzam2021single}. \\
Here we study lasing in a plasmonic square lattice where we create a periodic supercell by removing particles at designated positions. We perform lasing experiments by combining the array with a solution of dye molecules. Under optical pumping we observe multiple lasing peaks that emerge at different energies and non-zero wavevectors. 
To understand the interplay between the two periods in the system (underlying square and the supercell periods) we calculate the Empty Lattice Approximation (ELA) from the geometric structure factor and free photon dispersion~\cite{schokker2016lasing}.
We can see additional modes enabled by the much longer supercell period which would otherwise be suppressed in the original square lattice.
These new modes can be seen to exist at high-symmetry points of the Brillouin Zone (BZ) as defined by the supercell: 74th $\Gamma$- and 106th $X$-point if all possible modes are considered. While these modes are expected to exist in any square lattice matching the period of the supercell, by changing the positions of the particles in the unit-cell we can exhibit some control over the modes. This type of supercell and theoretical framework provides a new platform for designing multimode lasing systems.

\section*{Experiments}

\begin{figure}
    \centering
    \includegraphics[width=\textwidth]{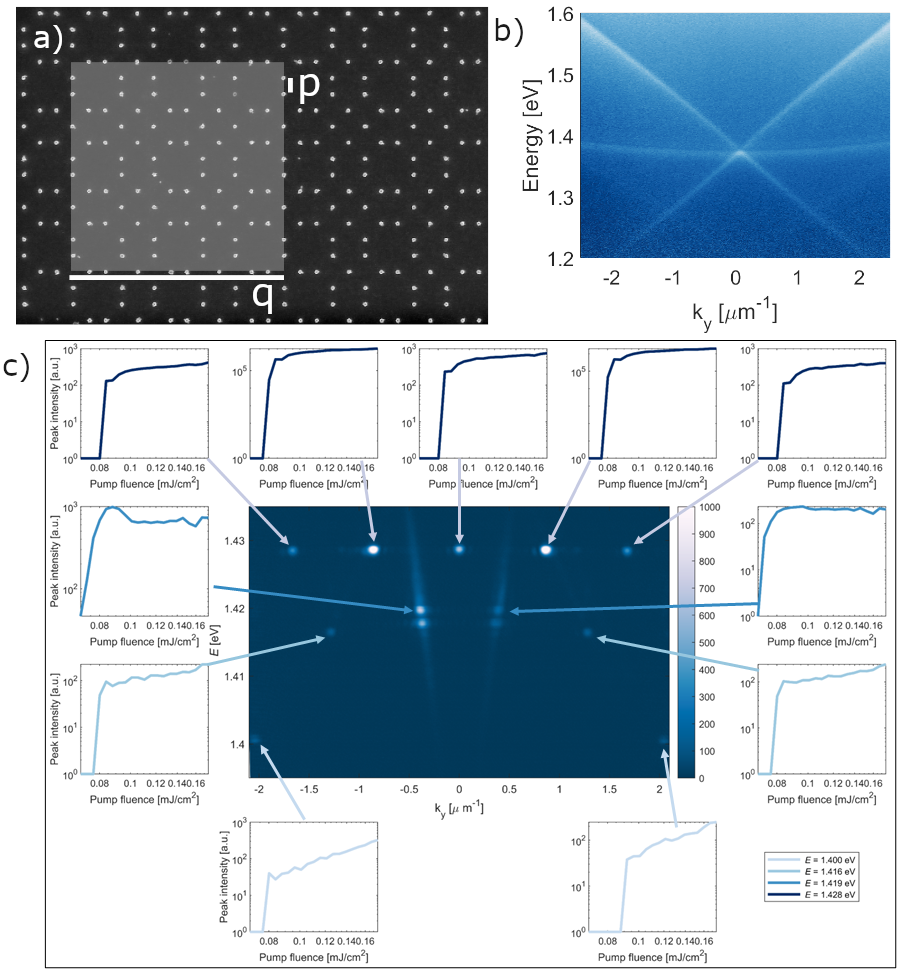}
    \caption{Lasing experiment in a supercell nanoparticle array. a) SEM image of the array studied. The array is based on a square array structure (period $p = 596$ nm) where specific lattice sites are kept empty. The giant unit cell is highlighted ($q$ is the unit cell size). b) Measured transmission spectrum and c) results of the lasing experiments. For the experiments in c), the array is combined with organic dye molecules (IR 140, 10 mM). Multiple lasing peaks are visible at different energies and $k_{y}$ (see central plot). The threshold behaviour of the lasing peaks is shown as indicated by the arrows.} 
    \label{fig:fig_1_experiments}
\end{figure}
We study a system based on a nanoparticle lattice with a square symmetry, however, part of the sites of the square array are left empty. This creates a unit cell of 13 $\times$ 13 sites which is repeated over the whole array. An SEM image of the array is shown in Fig. \ref{fig:fig_1_experiments} a). Details of the sample fabrication and the measurement setup are given in the Supporting Information, including Fig.~S1. The period of the square array is $p = 596$ nm and hence, the period of the supercell is $q = 7748$ nm. The gold nanoparticles have a diameter of 120 nm and a height of 50 nm. A transmission measurement of the array is shown in Fig. \ref{fig:fig_1_experiments} b), where the dispersive surface lattice resonances (SLRs) are clearly visible. The SLRs correspond to the underlying square array with the $\Gamma$-point located at $k_{y}=0$, $E$=1.37 eV. In transmission measurements, the finer features caused by the supercell are scarcely visible below the TM-branch. \\
We combine the nanoparticle array with a solution of organic dye molecules (IR 140, 10 mM) by sandwiching a droplet between the sample slide and a superstrate, and pump the system optically with a femtosecond laser (1 kHz repition rate, 800 nm central wavelength). With increasing pump fluence, a set of narrow peaks emerge as shown in Figure \ref{fig:fig_1_experiments} c). The corresponding threshold curves, i.e. emission intensity versus pump fluence are shown in the plots surrounding the spectrometer data recorded at a pump fluence of 0.1329 mJ/cm$^{2}$. At the highest energy of $1.428$ eV there are five peaks at $k_{y}= -1.662 \mu\mathrm{m}^{-1}$, $k_{y}= -0.852 \mu\mathrm{m}^{-1}$,  $k_{y}= 0 \mu\mathrm{m}^{-1}$, $k_{y}= 0.852 \mu\mathrm{m}^{-1}$, and $k_{y}= 1.677 \mu\mathrm{m}^{-1}$. There are four modes along the transverse electric (TE) SLRs at energies of $E = 1.419$ eV ($k_{y}= -0.387 \mu\mathrm{m}^{-1}$ and $k_{y}= 0.392 \mu\mathrm{m}^{-1}$) and $E = 1.417$ eV ($k_{y}= -0.370 \mu\mathrm{m}^{-1}$ and $k_{y}= 0.366 \mu\mathrm{m}^{-1}$). For clarity, only the threshold curves for the modes at the slightly higher energy are shown in Fig. \ref{fig:fig_1_experiments} c), the other modes are shown in the Supporting Information, Figure \ref{fig:threshold_lower_TESLR}. At energies of $E = 1.416$ eV are two peaks visible at $k_{y}= -1.281 \mu\mathrm{m}^{-1}$ and $k_{y}= 1.279 \mu\mathrm{m}^{-1}$. Lastly, there are two lasing peaks at $k_{y}= -2.046 \mu\mathrm{m}^{-1}$ and $k_{y}= 2.050 \mu\mathrm{m}^{-1}$ at an energy of $E = 1.400$ eV. Note that only the four lasing peaks that lie on the TE modes of the SLRs coindice with modes that can be seen in Fig. \ref{fig:fig_1_experiments} b), whereas none of the modes at other lasing peaks are visible in the transmission measurement.\\
Figure~S3 in the Supporting Information shows the dye emission with the lasing mode energies indicated. All of these modes coincide with the emission maximum. Interestingly, the $\Gamma$-point of the underlying square lattice ($E = 1.37$ eV)
is located at an energy where no lasing takes place. This suggests that the modes originating from the supercell experience more gain and/or have lower losses and are therefore more likely to lase. 
By changing the square lattice period, the dispersions of the arrays can be conveniently shifted with respect to the emission maximum of the dye. As a consequence, the modes that exhibit lasing can be tuned as shown in Figure~S4 in the Supporting Information.\\
\section*{Results and Discussion}
\begin{figure}
    \centering
    \includegraphics[width = \textwidth]{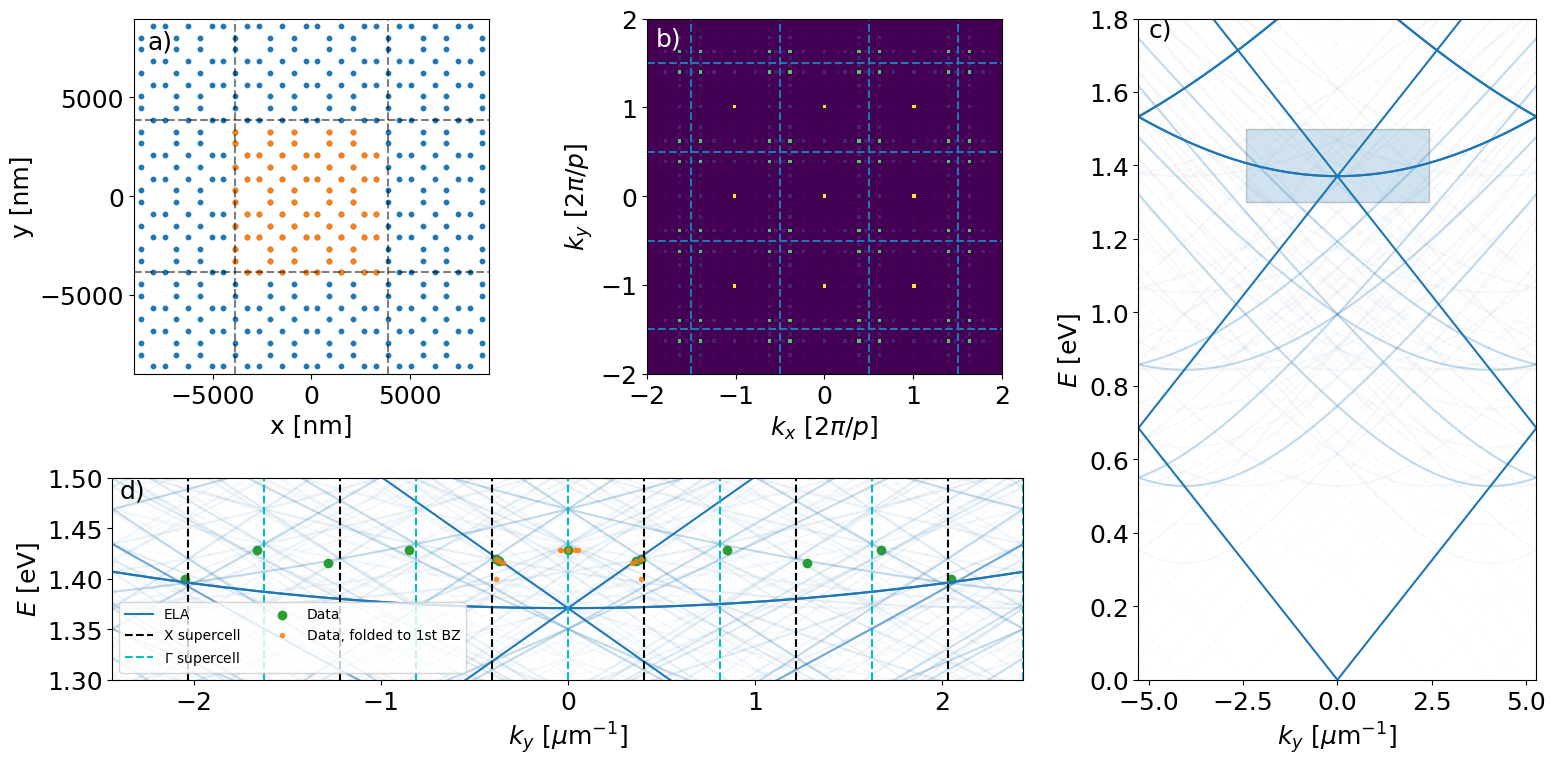}
    \caption{
    a) Schematic picture of the lattice with the unit cell (orange) and superlattice period highlighted.
    b) Structure factor calculated with equation \ref{eq:struct_fact} with dashed lines marking the reciprocal lattice as defined by the underlying square lattice.
    For clarity the values shown are limited to $[0,0.2]$.
    A close-up is shown in Figure~S6.
    c) Weighed empty-lattice approximation calculated from the structure factor. 
    Line color correlates with the ELA weight.
    d) Close-up of the dispersion relation close to the experimentally measured region (shaded area in c)).
    The weights have been increased to make the fine-structure more visible.
    Measured lasing peaks (green dots) are found to exist at crossings of at least two bands.
    When folded into the first BZ as defined by the superlattice (orange dots) period we find that lasing peaks at the same energy actually correspond to the same high-symmetry point.
    For visual clarity the $\Gamma$ and $X$-points of the supercell BZ have been highlighted in cyan and black dashed lines.
    }
    \label{fig:theoryfig}
\end{figure}

In Figure \ref{fig:theoryfig} the real space pattern of the nanoparticle array is shown (a) along with its geometric structure factor $S(\mathbf{k})$(b).
The structure factor describes the scattering properties of the lattice for any given wave-vector $\mathbf{k}$ and can be interpreted as a measure of constructive interference along that scattering direction.
In a typical square lattice $S(\mathbf{k})$ has peaks of equal magnitude at reciprocal lattice sites, i.e. at the centres of the Brillouin Zone (BZ). 
Removing particles from the array removes some of the destructive interference in the system allowing for new scattering directions to occur.
The amplitude of these new scatterings depends on the number and positions of the particles removed.
If particles are removed periodically, the system becomes periodic with a supercell period and has a new, smaller BZ.
However, these new BZ:s are not made equal as their properties are dependent on the way particles are removed.
Let us denote the initial square lattice period and the supercell period as $p$ and $q$ respectively, with associated reciprocal lattice vectors of magnitude $a$ and $b$.
For a periodic structure with a multi-particle unit cell, the normalized structure factor is
\begin{equation}
    S(\mathbf{k}) = \frac{1}{N^2}\sum_{ij} e^{i\mathbf{k}\cdot(\mathbf{r}_i - \mathbf{r}_j)}
                = \frac{1}{N_u^2}\sum_{i'j'}e^{i\mathbf{k}\cdot(\mathbf{r}_{i'} - \mathbf{r}_{j'})} \cdot 
                  \underbrace{\frac{1}{N_\alpha^2}\sum_{\alpha\beta}e^{i\mathbf{k}\cdot(\mathbf{q_\alpha - q_\beta})}}_{\delta(\mathbf{k} - m_1 \mathbf{b_1} - m_2\mathbf{b_2})},
                  \label{eq:struct_fact}
\end{equation}
where $N_u, N_\alpha$ are the number of particles in a unit cell and number of unit cells, $i,j$ and $i',j'$ the indices of particles in the lattice and in a unit cell, $\mathbf{q_\alpha, q_\beta}$ the positions of unit-cells, $\mathbf{b_1}$,$\mathbf{b_2}$ reciprocal lattice vectors and $m_1,m_2$ integers.
As can be seen in Figure \ref{fig:theoryfig} b), the system retains information from both periods.
In addition to the original scattering peaks (yellow dots at the $\Gamma$-points of the square lattice) a new set of secondary peaks appears at the $\Gamma$-points of the supercell lattice, the strongest of which exist near the corners of the initial BZ.
The new system retains information from the original period $p$ as the $S(\mathbf{k})$ can be seen to repeat with a period of $a = 2\pi/p$.


The first approximation for the band-structure is given by the empty-lattice approximation (ELA) which is obtained by taking the convolution of geometric structure factor $S(\mathbf{k})$ with the in-plane free-photon dispersion $|\mathbf{k}| = \sqrt{k_x^2 + k_y^2} = nE/(\hbar c)$ \cite{schokker2016lasing}: $\int d\mathbf{q} S(\mathbf{q})\delta(|\mathbf{q - k}| - nE/(\hbar c) )$.
This corresponds to placing light cones at the peaks of the structure factor (in case of square (super) lattices the $\Gamma$-points of each Brillouin zone) and weighing them by the value of the structure-factor. The results are shown in Figure \ref{fig:theoryfig} c).
As $S(\mathbf{k})$ indicates the amount of constructive interference, the weight of the light-cone correlates with the strength of the mode.
In addition to the typical ELA-dispersions of a square lattice, additional weaker dispersions emerge from the $\Gamma$-points of the supercell.
The experimentally measured dispersion overlayed with the calculated ELA is shown in Figure~S5 in the Supporting Information showing a good agreement with the theoretical model and measurements. 
By comparing the measurements to the ELA-dispersions, we find that the lasing peaks exist slightly below crossings of two or more bands at the high-symmetry points of the new BZ as is shown in figure \ref{fig:theoryfig} d). 
Due to the large size of the supercell, the new BZ is small and can be repeated multiple times in the measured range.
The modes at equal energy are separated by a multiple of $k_y = b$, so when folded back to the first BZ they can be seen to correspond to the same high-symmetry points. It is unclear whether in the experiments there exist even more lasing peaks at the same energy as we are limited in $k_y$ by the optics, i.e. the numerical aperture of the objective, (see Supporting Information Figure~S1). 
Interestingly, some of the bands in the band-structure in Figures \ref{fig:theoryfig} c) and d) have different slopes than the TE and TM modes originating from the underlying square lattice (the lines with the highest weight). These modes with different slopes cannot be categorized as purely transverse electric/magnetic (TE/TM) and the modes observed in the measurements coincide with such bands.This implies that the lasing modes are not purely TE or TM polarized.

\begin{figure}[htbp!]
    \centering
    \includegraphics[width=\textwidth]{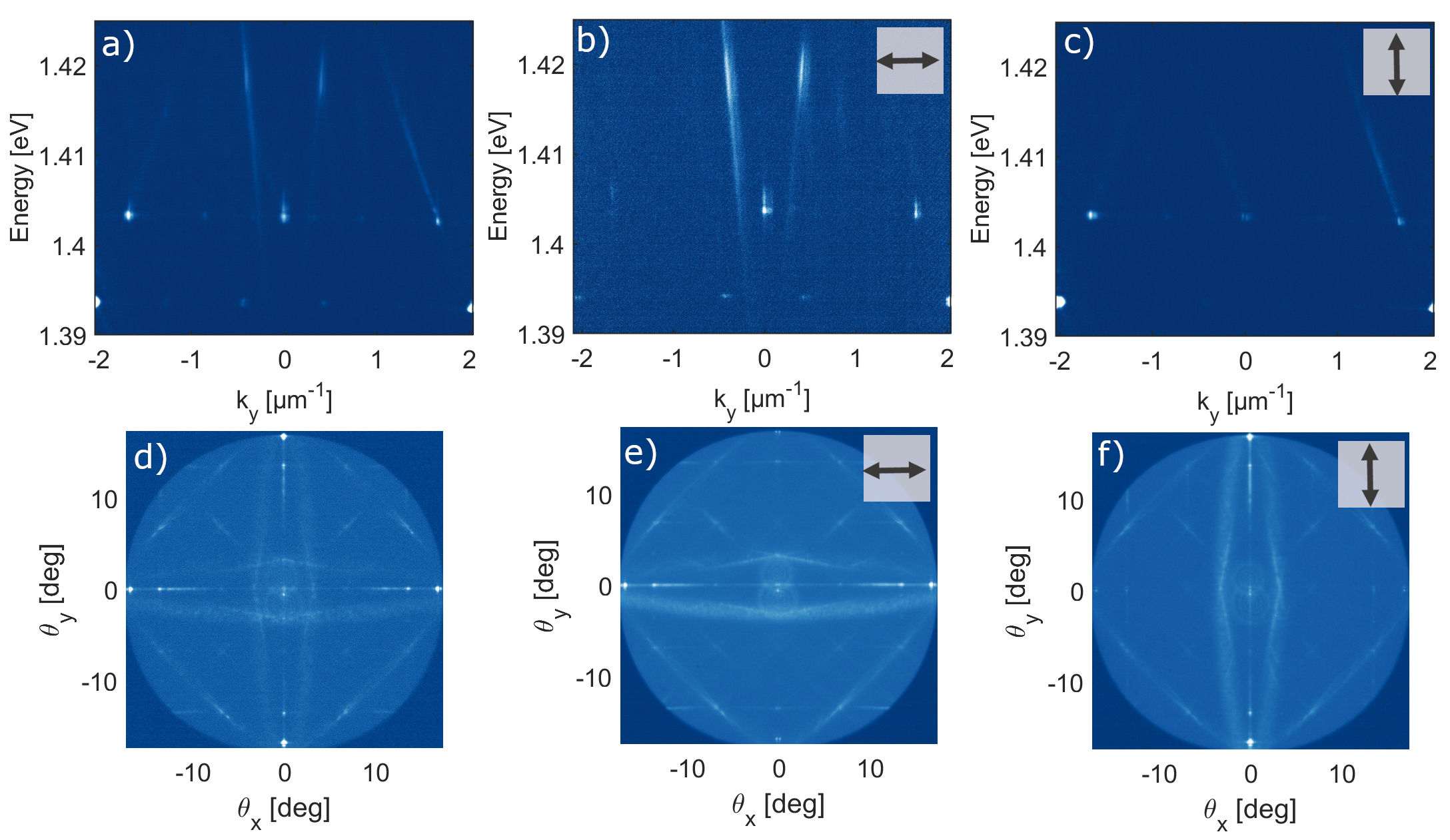}
    \caption{Polarization characteristics of the lasing modes in a large supercell array. The square lattice period is $p = 596$ nm and the edge length of the array is 240 $\mu$m. Angle-resolved spectra for a) no polarizer applied, b) horizontally, and c) vertically oriented polarizer. Full momentum space images in logarithmic scale for d) no polarizer applied, e) horizontally, and f) vertically oriented polarizer.} 
    \label{fig:pol_lasing}
\end{figure}

\noindent 
To verify the hybrid TE/TM nature of the lasing states, we experimentally studied an array with the same lattice parameters ($p = 596$ nm and $q = 13p$), however, the edge length is now $240$ $\mu$m. We combined the nanoparticle array with a reservoir of dye molecules with the same concentration of 10 mM as in the previous measurements, leading to increased total amount of molecules. The larger array as well as the increased gain caused by the higher amount of molecules lead to a stronger signal. This is needed as we added a polarizer into the detection path which decreases the amount of measurable light. We collected angle-resolved spectra as well as full momentum space images as is shown in Figure \ref{fig:pol_lasing}. Here, a vertical orientation refers to the axis of the polarizer oriented parallel $k_{x} = 0^{\circ}$ and a horizontal to the axis of the polarizer oriented parallel to $k_{y}=0^{\circ}$.\\
Changing the size of the array changes the lasing spectrum. In total, five lasing peaks are clearly observable in Figure \ref{fig:pol_lasing} a) with additional four peaks with less intensity. The peaks at the higher energy ($E = 1.403$ eV) are at wavevectors $k_{y} = -1.680, -0.859,  0,  0.845,$ and $1.661$ $\mu$m$^{-6}$ which correspond directly to the wavevectors of the highest energy mode ($E = 1.428$ eV) of the smaller array shown in Figure \ref{fig:fig_1_experiments} c), and based on the weights of the ELA we conclude that the lasing mode is now the 71st $\Gamma$-point.
The peaks at the lower energy in Figure \ref{fig:pol_lasing} a) ($ E = 1.393$ eV) occur at wavevectors $ -2.047, -0.443, 0.431, $ and $ 2.024$ $\mu$m$^{-6}$, where the larger $k_{y}$ directly correspond to the $X$-point lasing peaks in Figure \ref{fig:fig_1_experiments} c). The peaks at the smaller $k_{y}$ are the $X$-points closest to $k_{y}$ in Figure \ref{fig:theoryfig} d). The shift in energy is most likely caused by the increased amount of dye molecules that leads to a shift in the refractive index.\\ 
The majority of these peaks are visible in the angle-resolved spectra with polarization filters applied, albeit with varying intensity. Further, the peaks at $k_{y} = \pm 0.8$ $\mu$m$^{-1}$ and $E = 1.403$ eV do not appear in the case where a horizontally oriented polarizer is applied (Figure \ref{fig:pol_lasing}) b), implying that these modes are TM polarized.\\
The full momentum space images in Figure \ref{fig:pol_lasing}, bottom row, show strong features along $\theta_{y/x}=0^{\circ}$ if a horizontally/vertically oriented polarizer is applied. These images are in logarithmic scale to make weaker features more visible. And indeed, although a polarizer is applied, the peaks along $\theta_{x/y}=0^{\circ}$ can still be distinguished with a horizontal/vertical polarization filter applied. Nevertheless, these peaks are much weaker in intensity as the others and although this implies the hybrid TE/TM nature of the modes, the modes are mainly TE or TM polarized.

For comparison, we simulated a superlattice-type version of our structure and found the results of such calculation to be in agreement with \cite{wang2017band}, as is shown in Supporting Information~S7. Since both structures have the same supercell periodicity, band crossings happen at same values of $\mathbf{k}$ and $E$. However, due to the differences in structure factors, these crossings have different weights, and thus different bands are expected to be responsible for lasing. The strongest secondary bands in the supercell lattice exist in the vicinity of the original square lattice modes while in our case the strongest secondary bands are found closer to the $M$-points. In this case, the observed crossings are not the results of these stronger bands as is evident in Figure \ref{fig:theoryfig} c) and d). Instead, we find the crossings to come from multiple colliding weak modes. While these modes exist for the superlattice structure as well, the relative weights of the modes can be estimated to be different, as is shown by the different weight of the lines in Figure \ref{fig:theoryfig} and ~S7. In fact, the relative strengths of these crossings can be seen to correlate with the mode brightness.

\section*{Conclusion}
We demonstrated multi-mode lasing in a plasmonic supercell lattice. We showed that by introducing a periodic supercell in a square lattice geometry, additional band edges are formed near high symmetry points of the supercell. These band edges enable lasing and we observed lasing in multiple modes. By calculating the empty lattice approximation based on the structure factor of the lattice design, we identified the lasing modes to be the 74th $\Gamma$- and 106th $X$-point of the supercell. Due to their higher order nature these lasing modes are not purely TE or TM polarized. By tuning the square lattice period with respect to the emission maximum of the gain medium we were able to select the lasing modes.
One advantage of the supercell approach compared to the superlattice approach is the relative size of the structure. In previous superlattices in which multimode lasing has been achieved, the individual arrays of the size of 10s of $\mu$m were arranged on a centimeter square ($10^{-4}$ m) scale \cite{wang2017band}. The supercell arrays presented in this work on the other hand provide multimode in arrays with a size of 115 $\mu$m x 115 $\mu$m ($10^{-8}$ m) and are therefore significantly smaller. 
The supercell design approach provides new possibilities to engineer band edges for multimode lasing for instance by distributing the particles within the supercell differently or by changing the supercell period. 

\section*{Acknowledgements}
We thank Grazia Salerno for valuable discussions on the calculation of the band structure based on the structure factor.\\
\textbf{Funding:} This work was supported by the Academy of Finland under Project No. 318937 (PROFI), 322002, the Academy of Finland Flagship Programme, Photonics Research and Innovation (PREIN), Project No. 320166 and 320167, and the Vilho, Yrjö and Kalle Väisälä Foundation. Part of the research was performed at the OtaNano Nanofab cleanroom (Micronova Nanofabrication Centre), supported by Aalto University. We acknowledge the computational resources provided by the Aalto Science-IT project. R.H. acknowledges financial support by the Finnish Foundation for Technology Promotion.\\
\textbf{Author contributions:} R.H. initiated the project and P.T. supervised it. R.H. fabricated the samples and performed the measurements. K.A. calculated the band structures. All authors discussed the results. R.H. and T.K.H. wrote the manuscript with input from all coauthors.\\
\textbf{Competing interests:} There are no competing financial interests.\\
\section*{Supporting Information}
The Supporting Information is available free of charge. Experimental Methods; Threshold curves of lower TE SLR lasing modes; Dye emission; Lasing experiments with different lattice periods; Measured dispersion and calculated ELA; Close-up of the structure factor; ELA of a superlattice. 
\bibliography{bibliography}

\begin{thebibliography}{10}

\bibitem{kravets2008extremely}
VG~Kravets, F~Schedin, and AN~Grigorenko.
\newblock Extremely narrow plasmon resonances based on diffraction coupling of
  localized plasmons in arrays of metallic nanoparticles.
\newblock {\em Physical {R}eview {L}etters}, 101(8):087403, 2008.

\bibitem{chu2008experimental}
Yizhuo Chu, Ethan Schonbrun, Tian Yang, and Kenneth~B Crozier.
\newblock Experimental observation of narrow surface plasmon resonances in gold
  nanoparticle arrays.
\newblock {\em Applied {P}hysics {L}etters}, 93(18):181108, 2008.

\bibitem{auguie2008collective}
Baptiste Augui{\'e} and William~L Barnes.
\newblock Collective resonances in gold nanoparticle arrays.
\newblock {\em Physical {R}eview {L}etters}, 101(14):143902, 2008.

\bibitem{giannini2011plasmonic}
Vincenzo Giannini, Antonio~I Fern{\'a}ndez-Dom{\'\i}nguez, Susannah~C Heck, and
  Stefan~A Maier.
\newblock Plasmonic nanoantennas: fundamentals and their use in controlling the
  radiative properties of nanoemitters.
\newblock {\em Chemical reviews}, 111(6):3888--3912, 2011.

\bibitem{rodriguez2011coupling}
Said Rahimzadeh~Kalaleh Rodriguez, Aimi Abass, Bj{\"o}rn Maes, Olaf~TA Janssen,
  Gabriele Vecchi, and J~G{\'o}mez Rivas.
\newblock Coupling bright and dark plasmonic lattice resonances.
\newblock {\em Physical Review X}, 1(2):021019, 2011.

\bibitem{humphrey2014plasmonic}
Alastair~D Humphrey and William~L Barnes.
\newblock Plasmonic surface lattice resonances on arrays of different lattice
  symmetry.
\newblock {\em Physical {R}eview {B}}, 90(7):075404, 2014.

\bibitem{ross2016optical}
Michael~B Ross, Chad~A Mirkin, and George~C Schatz.
\newblock Optical properties of one-, two-, and three-dimensional arrays of
  plasmonic nanostructures.
\newblock {\em The journal of physical chemistry C}, 120(2):816--830, 2016.

\bibitem{vakevainen2014plasmonic}
A~I V{\"a}kev{\"a}inen, RJ~Moerland, HT~Rekola, A-P Eskelinen, J-P Martikainen,
  D-H Kim, and P~T{\"o}rm{\"a}.
\newblock Plasmonic surface lattice resonances at the strong coupling regime.
\newblock {\em Nano {L}etters}, 14(4):1721--1727, 2014.

\bibitem{bozhevolnyi2017quantum}
Sergey~I Bozhevolnyi, Luis Martin-Moreno, and Francisco Garcia-Vidal.
\newblock {\em Quantum {P}lasmonics}.
\newblock Springer, 2017.

\bibitem{hakala2018bose}
Tommi~K Hakala, Antti~J Moilanen, Aaro~I V{\"a}kev{\"a}inen, Rui Guo,
  Jani-Petri Martikainen, Konstantinos~S Daskalakis, Heikki~T Rekola, Aleksi
  Julku, and P{\"a}ivi T{\"o}rm{\"a}.
\newblock Bose--einstein condensation in a plasmonic lattice.
\newblock {\em Nature {P}hysics}, 14(7):739--744, 2018.

\bibitem{de2018interaction}
Milena De~Giorgi, Mohammad Ramezani, Francesco Todisco, Alexei Halpin, Davide
  Caputo, Antonio Fieramosca, Jaime Gomez-Rivas, and Daniele Sanvitto.
\newblock Interaction and coherence of a plasmon--exciton polariton condensate.
\newblock {\em ACS Photonics}, 5(9):3666--3672, 2018.

\bibitem{wang2018rich}
Weijia Wang, Mohammad Ramezani, Aaro~I V{\"a}kev{\"a}inen, P{\"a}ivi
  T{\"o}rm{\"a}, Jaime~G{\'o}mez Rivas, and Teri~W Odom.
\newblock The rich photonic world of plasmonic nanoparticle arrays.
\newblock {\em Materials {T}oday}, 21(3):303--314, 2018.

\bibitem{vakevainen2020sub}
Aaro~I V{\"a}kev{\"a}inen, Antti~J Moilanen, Marek Ne{\v{c}}ada, Tommi~K
  Hakala, Konstantinos~S Daskalakis, and P{\"a}ivi T{\"o}rm{\"a}.
\newblock Sub-picosecond thermalization dynamics in condensation of strongly
  coupled lattice plasmons.
\newblock {\em Nature {C}ommunications}, 11(1):3139, 2020.

\bibitem{zhou2013lasing}
Wei Zhou, Montacer Dridi, Jae~Yong Suh, Chul~Hoon Kim, Dick~T Co, Michael~R
  Wasielewski, George~C Schatz, and Teri~W Odom.
\newblock Lasing action in strongly coupled plasmonic nanocavity arrays.
\newblock {\em Nature {N}anotechnology}, 8(7):506--511, 2013.

\bibitem{schokker2014lasing}
A~Hinke Schokker and A~Femius Koenderink.
\newblock Lasing at the band edges of plasmonic lattices.
\newblock {\em Physical {R}eview {B}}, 90(15):155452, 2014.

\bibitem{yang2015real}
Ankun Yang, Thang~B Hoang, Montacer Dridi, Claire Deeb, Maiken~H Mikkelsen,
  George~C Schatz, and Teri~W Odom.
\newblock Real-time tunable lasing from plasmonic nanocavity arrays.
\newblock {\em Nature {C}ommunications}, 6(1):6939, 2015.

\bibitem{ramezani2017plasmon}
Mohammad Ramezani, Alexei Halpin, Antonio~I Fern{\'a}ndez-Dom{\'\i}nguez,
  Johannes Feist, Said Rahimzadeh-Kalaleh Rodriguez, Francisco~J Garcia-Vidal,
  and Jaime~G{\'o}mez Rivas.
\newblock Plasmon-exciton-polariton lasing.
\newblock {\em Optica}, 4(1):31--37, 2017.

\bibitem{guo2019lasing}
Rui Guo, Marek Ne{\v{c}}ada, Tommi~K Hakala, Aaro~I V{\"a}kev{\"a}inen, and
  P{\"a}ivi T{\"o}rm{\"a}.
\newblock Lasing at k points of a honeycomb plasmonic lattice.
\newblock {\em Physical {R}eview {L}etters}, 122(1):013901, 2019.

\bibitem{juarez2022m}
Xitlali~G Juarez, Ran Li, Jun Guan, Thaddeus Reese, Richard~D Schaller, and
  Teri~W Odom.
\newblock M-point lasing in hexagonal and honeycomb plasmonic lattices.
\newblock {\em ACS {P}hotonics}, 9(1):52--58, 2022.

\bibitem{heilmann2022quasi}
Rebecca Heilmann, Grazia Salerno, Javier Cuerda, Tommi~K Hakala, and P{\"a}ivi
  T{\"o}rm{\"a}.
\newblock Quasi-{BIC} mode lasing in a quadrumer plasmonic lattice.
\newblock {\em ACS {P}hotonics}, 9(1):224--232, 2022.

\bibitem{mohamed2022controlling}
Sughra Mohamed, Jie Wang, Heikki Rekola, Janne Heikkinen, Benjamin Asamoah, Lei
  Shi, and Tommi~K Hakala.
\newblock Controlling topology and polarization state of lasing photonic bound
  states in continuum.
\newblock {\em Laser \& Photonics {R}eviews}, 16(7):2100574, 2022.

\bibitem{salerno2022loss}
Grazia Salerno, Rebecca Heilmann, Kristian Arjas, Kerttu Aronen, Jani-Petri
  Martikainen, and P{\"a}ivi T{\"o}rm{\"a}.
\newblock Loss-driven topological transitions in lasing.
\newblock {\em Physical {R}eview {L}etters}, 129(17):173901, 2022.

\bibitem{lu2014all}
Yu-Jung Lu, Chun-Yuan Wang, Jisun Kim, Hung-Ying Chen, Ming-Yen Lu, Yen-Chun
  Chen, Wen-Hao Chang, Lih-Juann Chen, Mark~I Stockman, Chih-Kang Shih, et~al.
\newblock All-color plasmonic nanolasers with ultralow thresholds: autotuning
  mechanism for single-mode lasing.
\newblock {\em Nano {L}etters}, 14(8):4381--4388, 2014.

\bibitem{liu2018laser}
Xiaoyan Liu, Suyu Yi, Xiaolin Zhou, Shuailong Zhang, Zhilai Fang, Zhi-Jun Qiu,
  Laigui Hu, Chunxiao Cong, Lirong Zheng, Ran Liu, et~al.
\newblock Laser-based white-light source for high-speed underwater wireless
  optical communication and high-efficiency underwater solid-state lighting.
\newblock {\em Optics {E}xpress}, 26(15):19259--19274, 2018.

\bibitem{zhao2019full}
Jinyang Zhao, Yongli Yan, Zhenhua Gao, Yuxiang Du, Haiyun Dong, Jiannian Yao,
  and Yong~Sheng Zhao.
\newblock Full-color laser displays based on organic printed microlaser arrays.
\newblock {\em Nature {C}ommunications}, 10(1):870, 2019.

\bibitem{guan2021plasmonic}
Jun Guan, Ran Li, Xitlali~G Juarez, Alexander~D Sample, Yi~Wang, George~C
  Schatz, and Teri~W Odom.
\newblock Plasmonic nanoparticle lattice devices for white-light lasing.
\newblock {\em Advanced {M}aterials}, page 2103262, 2021.

\bibitem{wang2015superlattice}
Danqing Wang, Ankun Yang, Alexander~J Hryn, George~C Schatz, and Teri~W Odom.
\newblock Superlattice plasmons in hierarchical {A}u nanoparticle arrays.
\newblock {\em ACS {P}hotonics}, 2(12):1789--1794, 2015.

\bibitem{wang2017band}
Danqing Wang, Ankun Yang, Weijia Wang, Yi~Hua, Richard~D Schaller, George~C
  Schatz, and Teri~W Odom.
\newblock Band-edge engineering for controlled multi-modal nanolasing in
  plasmonic superlattices.
\newblock {\em Nature {N}anotechnology}, 12(9):889--894, 2017.

\bibitem{wang2019manipulating}
Danqing Wang, Jun Guan, Jingtian Hu, Marc~R Bourgeois, and Teri~W Odom.
\newblock Manipulating light--matter interactions in plasmonic nanoparticle
  lattices.
\newblock {\em Accounts of {C}hemical {R}esearch}, 52(11):2997--3007, 2019.

\bibitem{guo2019plasmon}
Ke~Guo, Sachin Kasture, and A~Femius Koenderink.
\newblock Plasmon antenna array “patchwork” lasers—towards low etendue,
  speckle free light sources.
\newblock {\em OSA {C}ontinuum}, 2(6):1982--1997, 2019.

\bibitem{cuartero2020super}
Alvaro Cuartero-Gonz{\'a}lez, Stephen Sanders, Lauren Zundel, Antonio~I
  Fern{\'a}ndez-Dom{\'\i}nguez, and Alejandro Manjavacas.
\newblock Super-and subradiant lattice resonances in bipartite nanoparticle
  arrays.
\newblock {\em ACS {N}ano}, 14(9):11876--11887, 2020.

\bibitem{lim2022fourier}
Theng-Loo Lim, Yaswant Vaddi, M~Saad Bin-Alam, Lin Cheng, Rasoul Alaee, Jeremy
  Upham, Mikko~J Huttunen, Ksenia Dolgaleva, Orad Reshef, and Robert~W Boyd.
\newblock Fourier-engineered plasmonic lattice resonances.
\newblock {\em ACS {N}ano}, 16(4):5696--5703, 2022.

\bibitem{zundel2021lattice}
Lauren Zundel, Asher May, and Alejandro Manjavacas.
\newblock Lattice resonances induced by periodic vacancies in arrays of
  nanoparticles.
\newblock {\em ACS {P}hotonics}, 8(1):360--368, 2021.

\bibitem{schokker2016lasing}
A~Hinke Schokker and A~Femius Koenderink.
\newblock Lasing in quasi-periodic and aperiodic plasmon lattices.
\newblock {\em Optica}, 3(7):686--693, 2016.

\bibitem{knudson2019polarization}
Michael~P Knudson, Ran Li, Danqing Wang, Weijia Wang, Richard~D Schaller, and
  Teri~W Odom.
\newblock Polarization-dependent lasing behavior from low-symmetry nanocavity
  arrays.
\newblock {\em ACS {N}ano}, 13(7):7435--7441, 2019.

\bibitem{guan2021identification}
Jun Guan, Marc~R Bourgeois, Ran Li, Jingtian Hu, Richard~D Schaller, George~C
  Schatz, and Teri~W Odom.
\newblock Identification of {B}rillouin {Z}ones by in-plane lasing from
  light-cone surface lattice resonances.
\newblock {\em ACS {N}ano}, 15(3):5567--5573, 2021.

\bibitem{tan2022lasing}
Max~JH Tan, Jeong-Eun Park, Francisco Freire-Fern{\'a}ndez, Jun Guan, Xitlali~G
  Juarez, and Teri~W Odom.
\newblock Lasing action from quasi-propagating modes.
\newblock {\em Advanced {M}aterials}, 34(34):2203999, 2022.

\bibitem{lin2020chip}
Ronghui Lin, Valerio Mazzone, Nasir Alfaraj, Jianping Liu, Xiaohang Li, and
  Andrea Fratalocchi.
\newblock On-chip hyperuniform lasers for controllable transitions in
  disordered systems.
\newblock {\em Laser \& Photonics {R}eviews}, 14(2):1800296, 2020.

\bibitem{kim2020multipolar}
Ha-Reem Kim, Min-Soo Hwang, Daria Smirnova, Kwang-Yong Jeong, Yuri Kivshar, and
  Hong-Gyu Park.
\newblock Multipolar lasing modes from topological corner states.
\newblock {\em Nature {C}ommunications}, 11(1):5758, 2020.

\bibitem{gong2020topological}
Yongkang Gong, Stephan Wong, Anthony~J Bennett, Diana~L Huffaker, and Sang~Soon
  Oh.
\newblock Topological insulator laser using valley-hall photonic crystals.
\newblock {\em ACS {P}hotonics}, 7(8):2089--2097, 2020.

\bibitem{zhai2023multimode}
Zhenshan Zhai, Zhuang Li, Yixuan Du, Xin Gan, Linye He, Xiaotian Zhang, Yufeng
  Zhou, Jun Guan, Yangjian Cai, and Xianyu Ao.
\newblock Multimode vortex lasing from dye--{T}i{O}2 lattices via bound states
  in the continuum.
\newblock {\em ACS {P}hotonics}, 10(2):437--446, 2023.

\bibitem{azzam2021single}
Shaimaa~I Azzam, Krishnakali Chaudhuri, Alexei Lagutchev, Zubin Jacob, Young~L
  Kim, Vladimir~M Shalaev, Alexandra Boltasseva, and Alexander~V Kildishev.
\newblock Single and multi-mode directional lasing from arrays of dielectric
  nanoresonators.
\newblock {\em Laser \& Photonics {R}eviews}, 15(3):2000411, 2021.

\end{thebibliography}
\bibliographystyle{unsrt}
\end{document}


\maketitle
\setcounter{equation}{0}
\setcounter{figure}{0}
\setcounter{table}{0}
\setcounter{page}{1}
\setcounter{section}{0}
\makeatletter
\renewcommand{\theequation}{S\arabic{equation}}
\renewcommand{\thefigure}{S\arabic{figure}}
\renewcommand{\thesection}{S-\Roman{section}}

\section*{Experimental Methods}
\subsection*{Fabrication}
The arrays are fabricated on a borosilicate substrate using electron beam lithography (EBL). First, a layer of Poly(methyl methacrylate) (PMMA, 4\% in anisol) is spincoated onto the substrate at 3000 rpm for two minutes, followed by baking for two minutes at 175$^{\circ}$C. A 10 nm thick layer of aluminium is evaporated on top as a conductive layer for the EBL. After the EBL exposure the aluminium layer is etched in a 1:1 mixture of AZ351B developer and de-ionized water. The PMMA is developed in a 1:3 mixture of methyl isobotyl ketone and isopropanol, followed by evaporation of 2 nm of titanium that acts as an adhesive layer as well as 50 nm of gold. Finally, the excess PMMA and metal are removed during a lift-off process in acetone.\\
The dye molecule solution is prepared by dissolving the IR 140 molecules in a 10 mM concentration into 1:2 mixture of dimethyl sulfoxide (DMSO) and benzyl alcohol (BA). The mixture of DMSO and BA ensures an index matching environment to the borosilicate substrate. In the lasing experiments as shown in Fig, \ref{fig:fig_1_experiments} c), a drop (10 $\mu$L) of the dye solutions is added onto the nanoparticle arrays and sealed by placing a second borosilicate substrate on top. In the polarization resolved measurements, the dye solution is injected into a cavity formed by a 1.6 mm thick press-to-seal silicone isolator ring sandwiched between the two borosilicate slides. In the transmission measurements the arrays are combined with index matching oil and sealed with a cover slip.\\
\subsection*{Measurement setup}
\begin{figure}
    \centering
    \includegraphics[width=\textwidth]{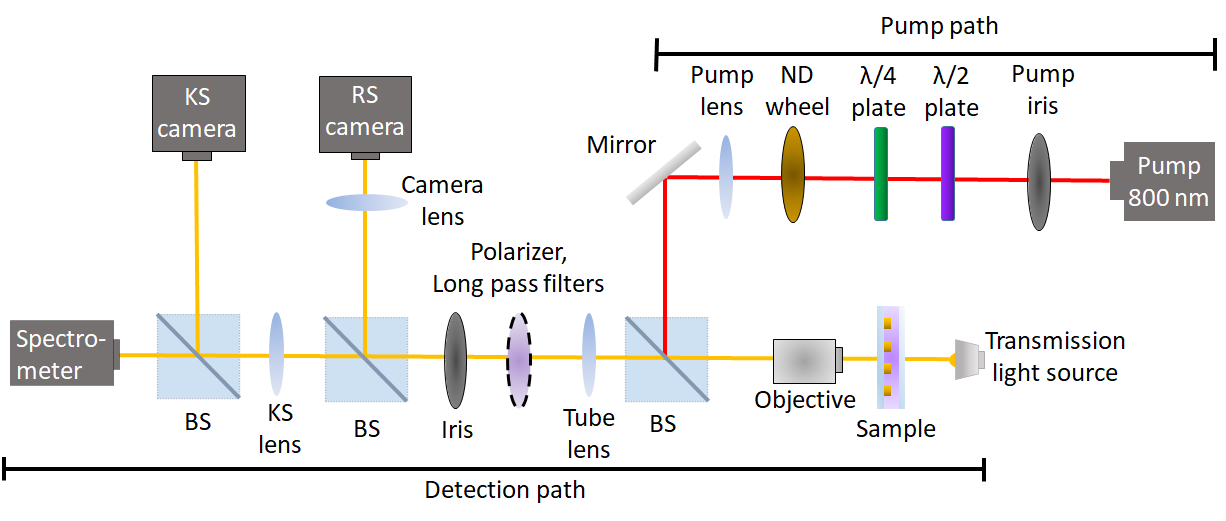}
    \caption{Experimental setup. Here, ND is neutral density, BS is beam splitter. This setup enables simultaneous measurements of the angle-resolved spectra (spectrometer CCD camera) and real space and momentum space images (CMOS cameras).}
    \label{fig:Setup}
\end{figure}

Figure \ref{fig:Setup} shows a schematic of the measurement setup. The transmission measurement is done by illuminating the sample with a white light halogen lamp. In the lasing experiments the system of dye molecules and the nanoparticle array is pumped with the pump laser (800 nm central wavelength, repetition rate 1 kHz). The pump covers the whole array area and is left circularly polarized. The measurements are taken as a function of pump fluence, where the pump fluence is controlled with the neutral density (ND) wheel. The signal of the array is collected with a 0.3 NA objective and the back focal plane of the objective is focused on the spectrometer slit. Each point on the slit corresponds to one angle on the sample and the CCD camera in the spectrometer collects the light so that the angular information is along one axis and the energy or wavelength information of the signal is along the other axis. Full real space and momentum space images can be recorded simultaneously with the spectrometer data using the CMOS cameras. Optional polarization filters and long pass filters can be added to the setup. During the lasing experiments a 850 nm long pass filter was used in order to cut off the signal of the pump laser.
\newpage
\section*{Threshold curves of lower TE SLR lasing modes}
 \begin{figure}[htbp!]
     \centering
     \includegraphics[width=\textwidth]{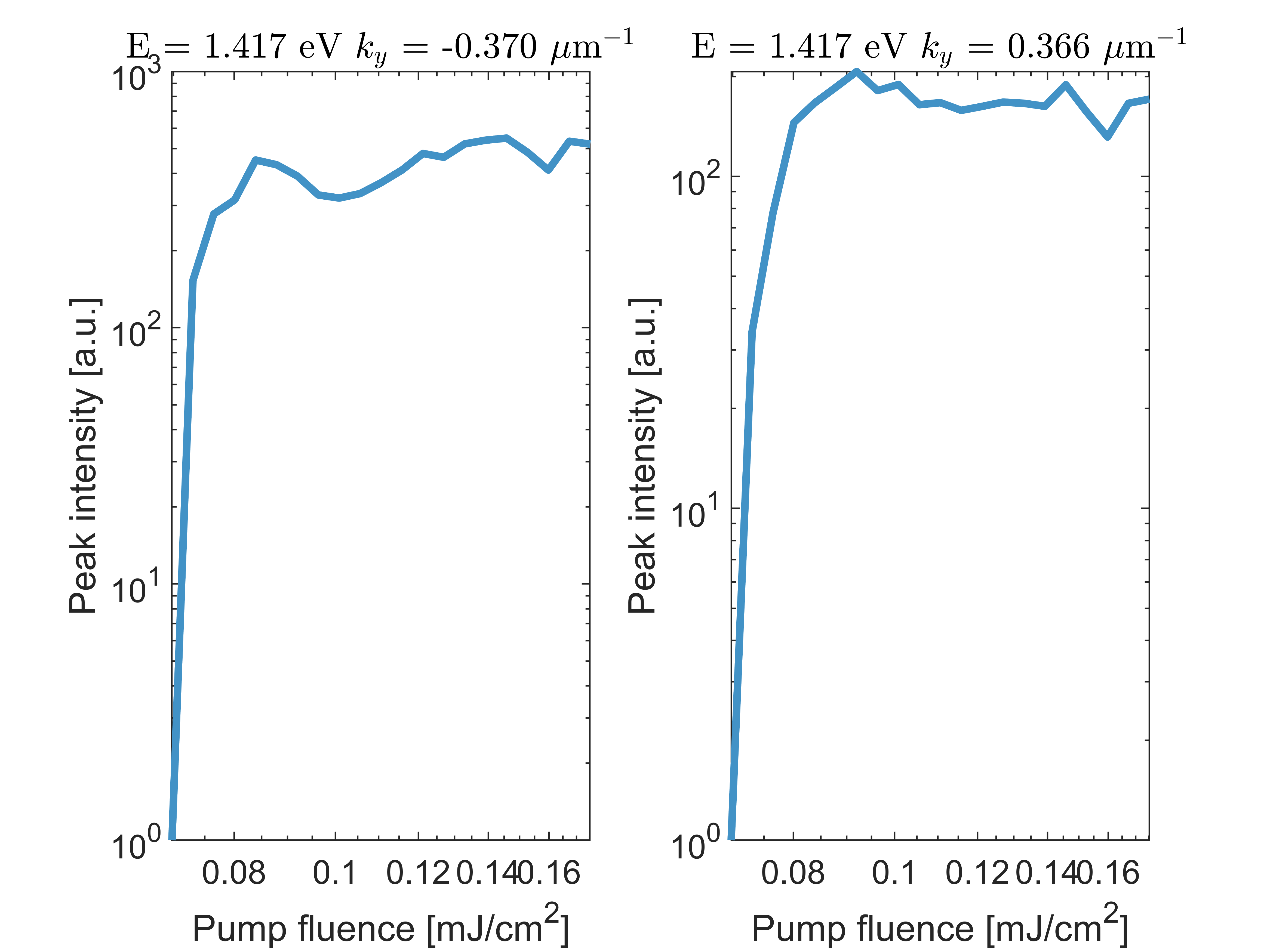}
     \caption{Threshold curves of lasing modes on TE SLRs at lower energy ($E = 1.417$ eV from main text Fig.~1 c)).}
     \label{fig:threshold_lower_TESLR}
 \end{figure}
\newpage
 \section*{Dye emission}
\begin{figure}[htbp!]
    \centering
    \includegraphics[width=0.9\textwidth]{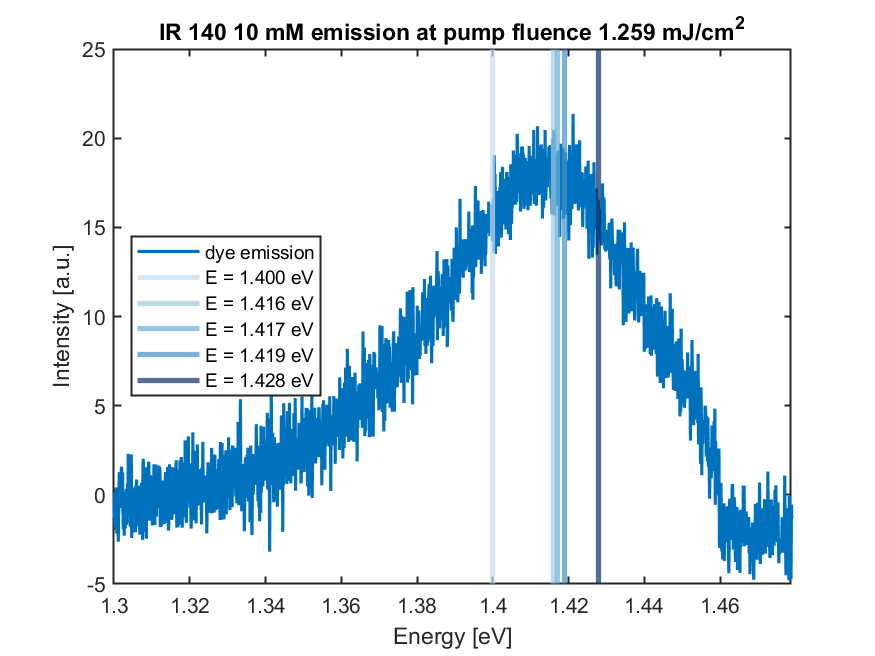}
    \caption{Emission spectrum of the dye (IR 140, 10 mM) pumped with a pump fluence of 1.259 mJ/cm$^{2}$. The lasing peak energies of the array ($p_{1}=596$ nm) in Fig.~1 are indicated. They clearly overlap with the emission maximum, explaining why these modes are lasing.}
    \label{fig:dye_emission}
\end{figure}
\newpage
\section*{Lasing experiments with different lattice periods}
\begin{figure}[htbp!]
    \centering
    \includegraphics[width=\textwidth]{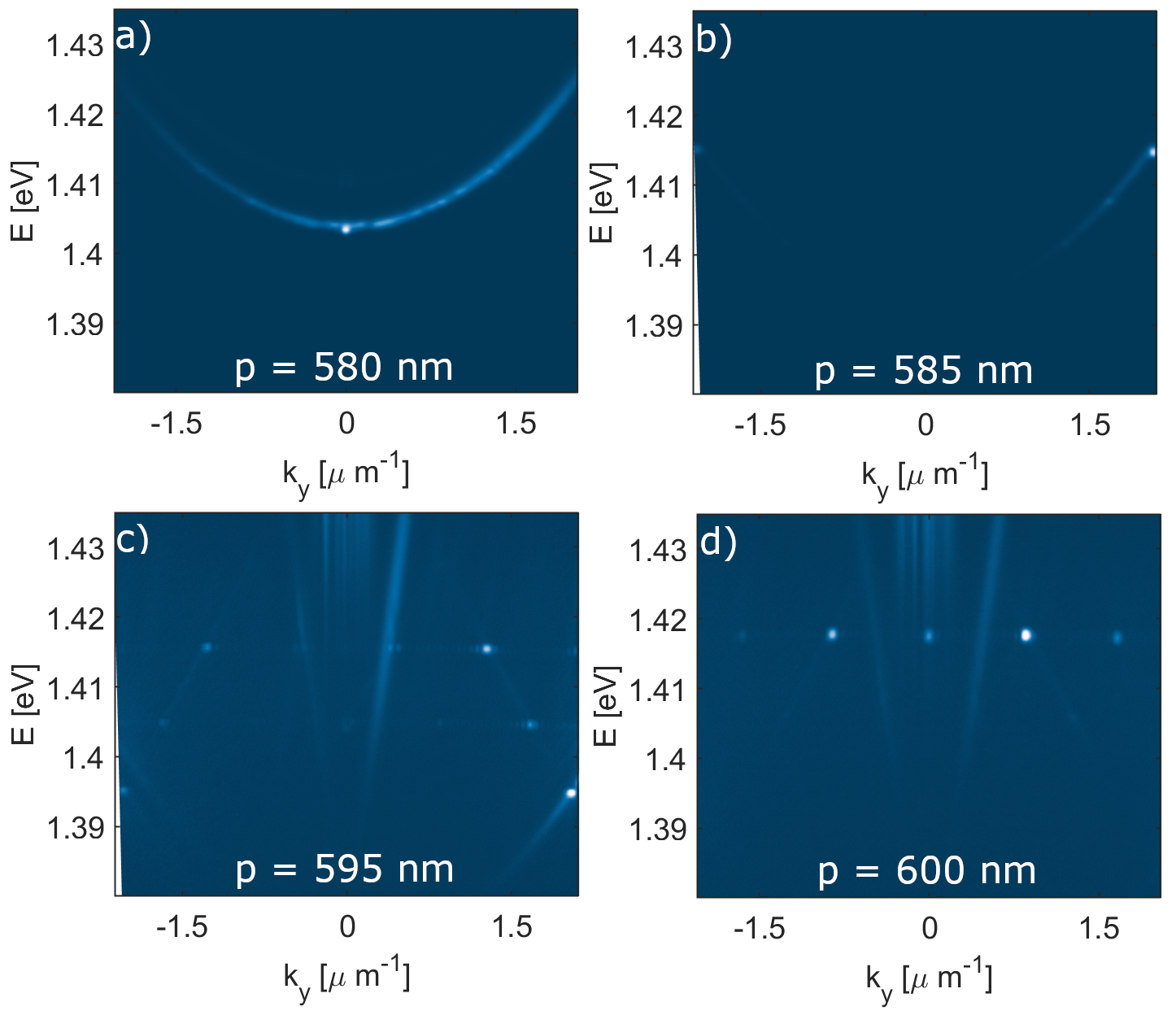}
    \caption{Angle-resolved lasing spectra for supercell arrays with different square lattice periods a) p = 580 nm, b) p = 585 nm, c) p =595 nm, and d) p =600 nm.}
    \label{fig:lasing_diff_p}
\end{figure}

We repeated the lasing experiments with different square lattice periods $p$ and the same supercell structure. The results for $p = 580$ nm, $590$ nm, $595$ nm, and $600$ nm are shown in Fig. \ref{fig:lasing_diff_p}. For the shortest period we observe $\Gamma$-point lasing at the $\Gamma$-point energy of the underlying square lattice. For increasing $p$ the dispersions are shifted to lower energies. For the next largest period $p = 585$ nm we observe lasing in only the two modes with the largest $k_{y}$, i.e. in the $X$-points. The next largest period $p =595$ nm enables lasing in the same modes as in the main text, where $p = 596$ nm. For an even larger period $p = 600$ nm, we observe lasing in only the five modes at the highest energy. Hence, by tuning the lattice period the modes are shifted with respect to the emission maximum of the dye and different modes can be chosen.

\newpage

\section*{Measured dispersion and calculated ELA}
\begin{figure}[htbp!]
    \centering
    \includegraphics[width=0.6\textwidth]{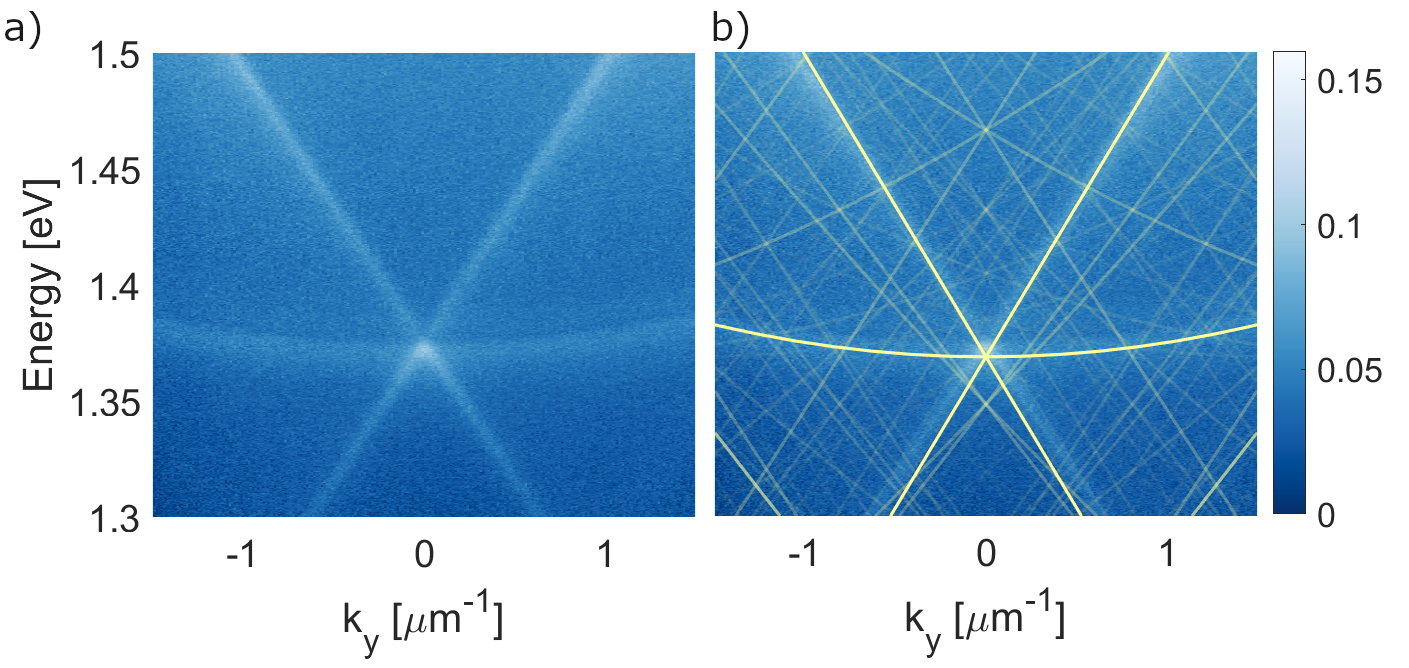}
    \caption{ a) Measured dispersion and b) measured dispersion with calculated ELA on plotted top. In the measured dispersion the TE and TM modes of the square lattice are the only modes visible. The overlap between the measured and calculated TE and TM modes is perfect. The other modes stemming from the supercell are too weak to be measured.}
    \label{fig:disp_ELA}
\end{figure}

\section*{Close-up of the structure-factor}

\begin{figure}[!ht]
    \centering
    \includegraphics[width=\textwidth]{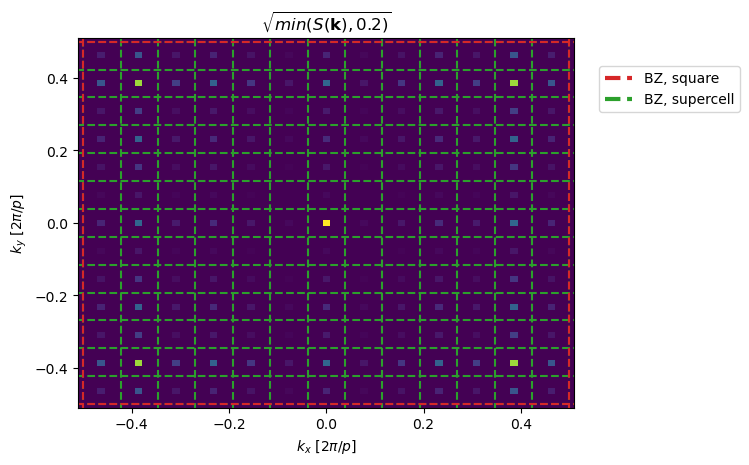}
    \caption{
    A close-up of the structure factor (fig~2 b)) with Brillouin Zones corresponding to supercell and square lattices highlighted.
    To highlight the finer structure inside each unit cell, the square root of the structure factor is plotted, and the values of the structure factor larger than 0.2 are replaced by that cut-off value.
    }
    \label{fig:BZ_zoom}
\end{figure}

\section*{Empty lattice approximation of a superlattice}

\begin{figure}[htbp!]
    \centering
    \includegraphics[width=\textwidth]{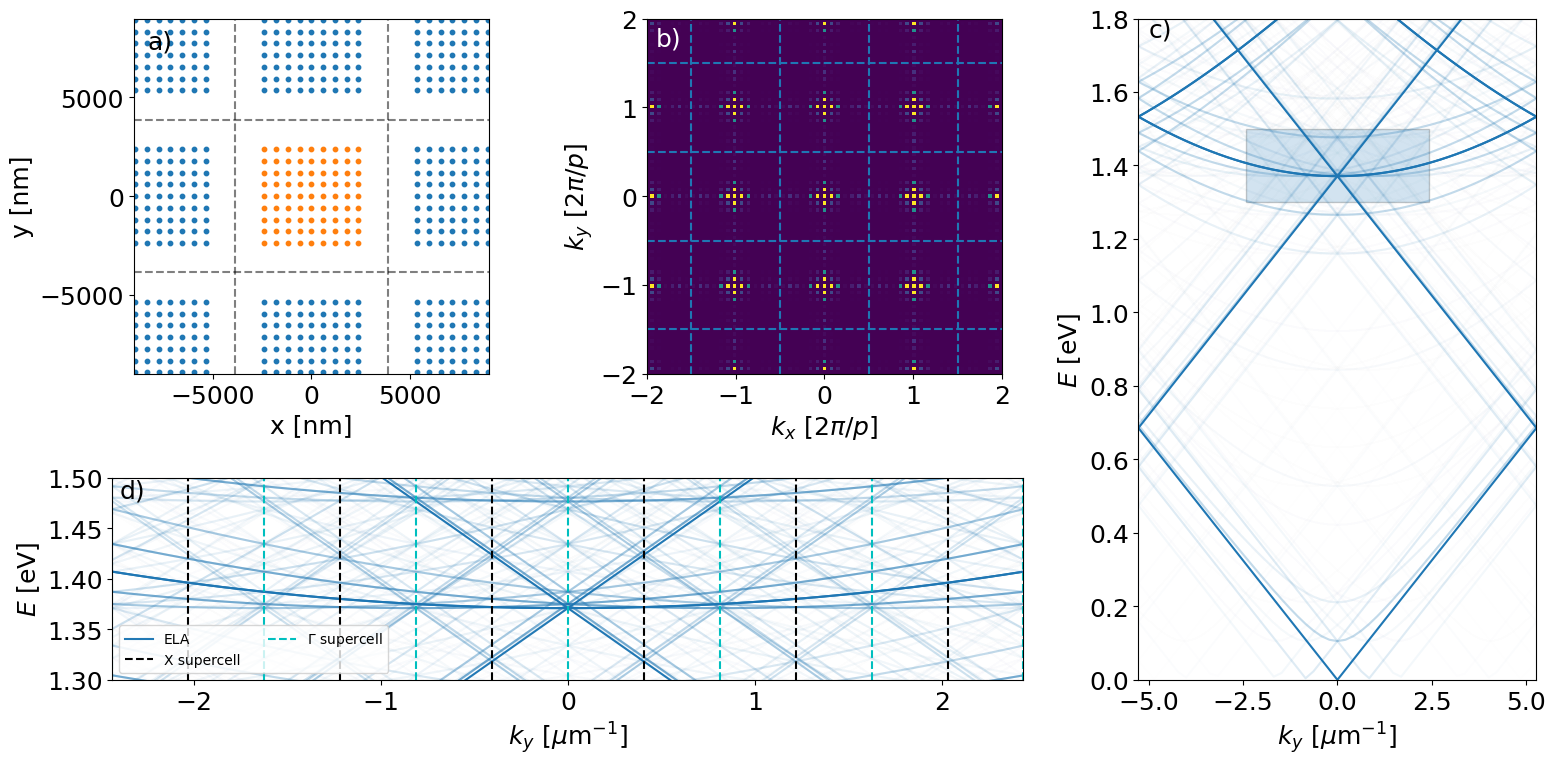}
    \caption{a) Schematic picture of the superlattice with the unit cell highlighted in orange. b) Calculated structure factor. c) Weighted empty-lattice approximation calculated from the structure factor. d) Close-up of the dispersion close to the experimentally measured region.}
    \label{fig:Superlattice}
\end{figure}

For comparison, a similar mode analysis as in the main text was done for a superlattice.
For this, the particles inside each supercell are rearranged into a square lattice with the period $p$. This leads to patches that are separated by the supercell period $q$. A schematic of this structure is shown in Figure \ref{fig:Superlattice} a). In order to create a full square one particle had to be added.\\
The structure factor of this structure is shown in Figure \ref{fig:Superlattice} b). This structure factor is clearly distinguishable from the structure factor of the supercell array presented in Figure~2 b) of the main text. The most striking difference is that the additional peaks of the structure factor are now located in the centre of the Brillouin Zones around the scattering maxima of the underlying square lattice. The peaks of the structure factor of the supercell lattice on the other hand are located along the edges of the Brillouin Zone.\\
Next, we calculate the band-structure of the superlattice structure following the method described in the main text. The resulting band-structure is shown in Figure \ref{fig:Superlattice} c) with a close up to the experimentally measured region in Figure \ref{fig:Superlattice} d). Due to the same periods of the square lattice and superlattice (or supercell in the main text), the band-crossings are at the same wavevectors and energies. However, the band-structures are clearly different from each other as different bands have a different weight. This is a direct result of the difference in the structure factor. As the structure factor is directly related to the scattering process in a structure, additional dispersive features arise, where constructive interference is allowed as shown by the structure factor. The weight of the dispersions is given by the the intensity and position of the peaks in the structure factor. Hence, by rearranging the particles within the supercell, we can control for which momenta constructive interference happens and hence change the weight in the dispersions. This weight difference is responsible for the differences in the proportional brightness of the lasing modes in our supercell lattice compared to the superlattice case of Ref.~[28].

